\pdfoutput=1
\documentclass[twocolumn,superscriptaddress,floatfix,preprintnumbers]{revtex4-1}
\usepackage{graphics,amssymb,amsmath,epsfig,color}
\usepackage{graphicx}

\begin{document}

\title{Supercoiling DNA locates mismatches}

\author{Andrew Dittmore$^1$, Sumitabha Brahmachari$^2$, Yasuhara Takagi$^1$, John F. Marko$^{2,3}$, and Keir C. Neuman}

\affiliation{
Laboratory of Single Molecule Biophysics, National Heart, Lung, and Blood Institute, National Institutes of Health, Bethesda, MD 20892\\
$^2$Department of Physics and Astronomy, Northwestern University,
Evanston IL 60208\\
$^3$Department of Molecular Biosciences, Northwestern University,
Evanston IL 60208
}

\date{\today}

\begin{abstract}

We present a method of detecting sequence defects by supercoiling DNA with magnetic tweezers. The method is sensitive to a single mismatched base pair in a DNA sequence of several thousand base pairs. We systematically compare DNA molecules with 0 to 16 adjacent mismatches at 1~$M$ monovalent salt and 3.5~pN force and show that, under these conditions, a single plectoneme forms and is stably pinned at the defect. We use these measurements to estimate the energy and degree of end-loop kinking at defects. From this, we calculate the relative probability of plectoneme pinning at the mismatch under physiologically relevant conditions. Based on this estimate, we propose that DNA supercoiling could contribute to mismatch and damage sensing {\it in vivo}.





\end{abstract}
\maketitle

With increasing imposed torsion, a thin elastic rod builds torque until it reaches a limit point, and then abruptly loops into a self-contacting, plied structure called a plectoneme \cite{purohit,benham}. Although in a macroscopic system this abrupt buckling is expected to occur at a site of elastic discontinuity \cite{goyal}, the analogous situation of a base-pair mismatch defect in a single DNA molecule is less clear: Thermal fluctuations could mask the effects of the defect in a DNA molecule of several thousand base-pairs. Alternatively, the bending energy decrease could overcome entropy, in which case the position of the defect would determine where the plectoneme forms. In addition to being a unique physical system in which to explore the role of defects in the torsional buckling of a fluctuating elastic rod, understanding how defects influence DNA supercoiling has potential biological implications. Whereas DNA supercoiling is known to be a major determinant of the large-scale organization of cellular DNA \cite{postow}, we propose that torque may also act globally to facilitate the detection of defect sites among millions of normal base pairs \cite{yang}.

Supercoiling potentially provides a mechanism of defect sensing by localizing a defect at a plectoneme tip, where the duplex will be further destabilized by bending stress. Previous work provides precedent for the idea of plectoneme pinning at DNA defects. Brutzer {\it et al.} showed overwound DNA buckles and forms a plectoneme that is pinned at a permanent kink \cite{brutzer}. Computations by Matek {\it et al.} \cite{matek} demonstrated that energy is minimized by the colocalization of plectoneme end-loops and regions of unpaired bases as ``tip bubbles." Ganji {\it et al.} \cite{ganji} used an intercalator based method for creating fluorescent plectonemes and showed a preference for plectoneme pinning at the location of a 10~bp mismatch.

Here we develop a framework to quantify how defects influence DNA supercoiling. By isolating individual plectonemes in DNA constrained by force and ionic screening, we find that even a single mismatch site specifies the location of a plectoneme. Statistical-mechanical calculations allow us to extend these results to physiologically relevant conditions.

\begin{figure}
    \centering
    \includegraphics[width=\columnwidth]
    {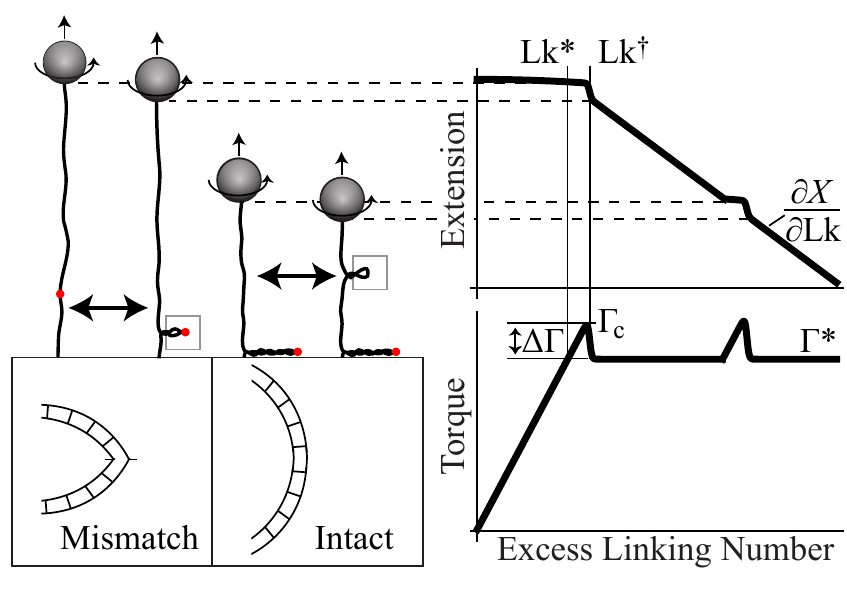}
    \caption{Method of locating a sequence defect in a single DNA molecule by supercoiling. We use magnetic tweezers to measure DNA extension at constant force while increasing the excess linking number ${\rm Lk}$ through applied turns of the tethered magnetic bead. For an asymmetrically positioned defect (red dot), two buckling transitions are observed. The first transition, at ${\rm Lk}^\dagger$ and critical torque $\Gamma_{\rm c}$, causes the torque to abruptly drop by an amount $\Delta\Gamma$ and produces a pinned plectoneme with the defect at its tip. Surface encounter of the initial plectoneme prevents it from lengthening and causes re-buckling of the DNA at a torque larger than $\Gamma_{\rm c}$. 
    The torque at ${\rm Lk}^*$ is assumed equal to the plateau torque $\Gamma^*$ that is independent of defects and determined from $|\partial{X}/\partial{{\rm Lk}}|$ after re-buckling \cite{clauvelin}.
   }
    \label{Fig1}
\end{figure}

Our experimental method is based on a magnetic tweezers supercoiling assay \cite{strick} and makes use of the abrupt, discontinuous buckling of DNA at the onset of plectoneme formation \cite{forth,brutzer} (Fig.~\ref{Fig1}). If a single pinned plectoneme forms at a defect positioned near a surface, it is prevented from lengthening as it impinges on the surface. This results in a second, abrupt buckling transition that occurs at double the distance of the defect from the surface. Thus the DNA extension at re-buckling reports the position of the defect.

\begin{figure*}
\label{Fig2}
\centering
\includegraphics{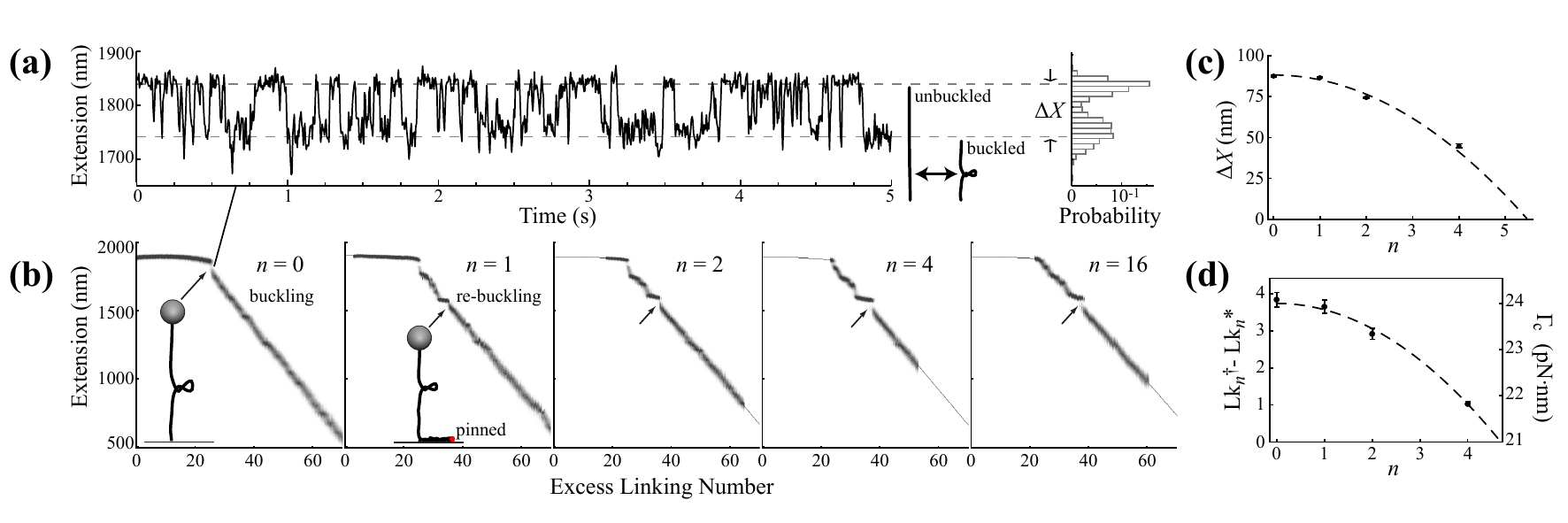}
\begin{minipage}{\textwidth}
\caption{Data from 6~kb DNA with $n=0$ to $16$ adjacent mismatches. (a) Example time-series recorded near linking number ${\rm Lk}={\rm Lk}^\dagger$ of fluctuations between unbuckled and buckled states, which differ in extension by $\Delta X$. (b) Data of extension vs excess linking number. All recorded points are plotted to show the variance and to highlight discrete changes. For $n=0$, only single buckling is observed (arrow); for $n>0$ re-buckling of intact DNA is observed (arrows) due to the surface encounter of the defect-pinned plectoneme. (c) The change in DNA extension upon buckling decreases quadratically with defect size, $n$. The dashed curve is the fit to $\Delta X_n=a_X-c_X\cdot n^2$, with $a_X=88.3\pm0.3$~nm and $c_X=2.94\pm0.08$~nm. (d) The linking number change (${\rm Lk}^\dagger-{\rm Lk}^*$, left axis) also decreases as $a_{{\rm Lk}}-c_{{\rm Lk}}\cdot n^2$, with best-fit parameters $a_{{\rm Lk}}=3.7\pm0.1$ and $c_{{\rm Lk}}=0.17\pm0.01$. This provides an estimate of the critical buckling torque (right axis); see text.}
\end{minipage}
\end{figure*}

\begin{figure}
    \centering
    \includegraphics[width=\columnwidth]{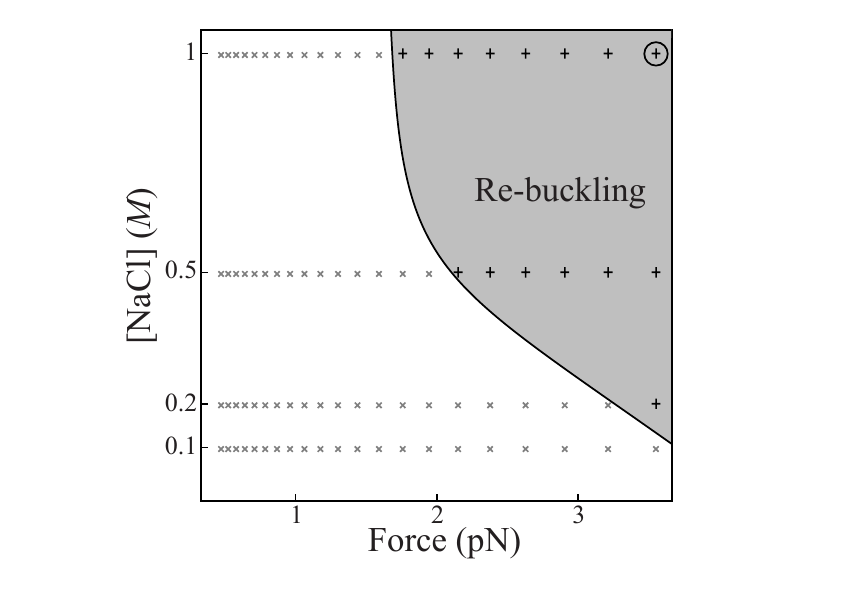}
    \caption{Experimental salt-force phase diagram of re-buckling for $n=2$. Re-buckling (+ points, grey shaded region) requires high force and ionic screening. The data in Fig.~2 were collected at 3.5~pN and $1~M$ salt (circled).}
    \label{Fig3}
\end{figure}

\begin{figure}
    \centering
    \includegraphics[width=\columnwidth]{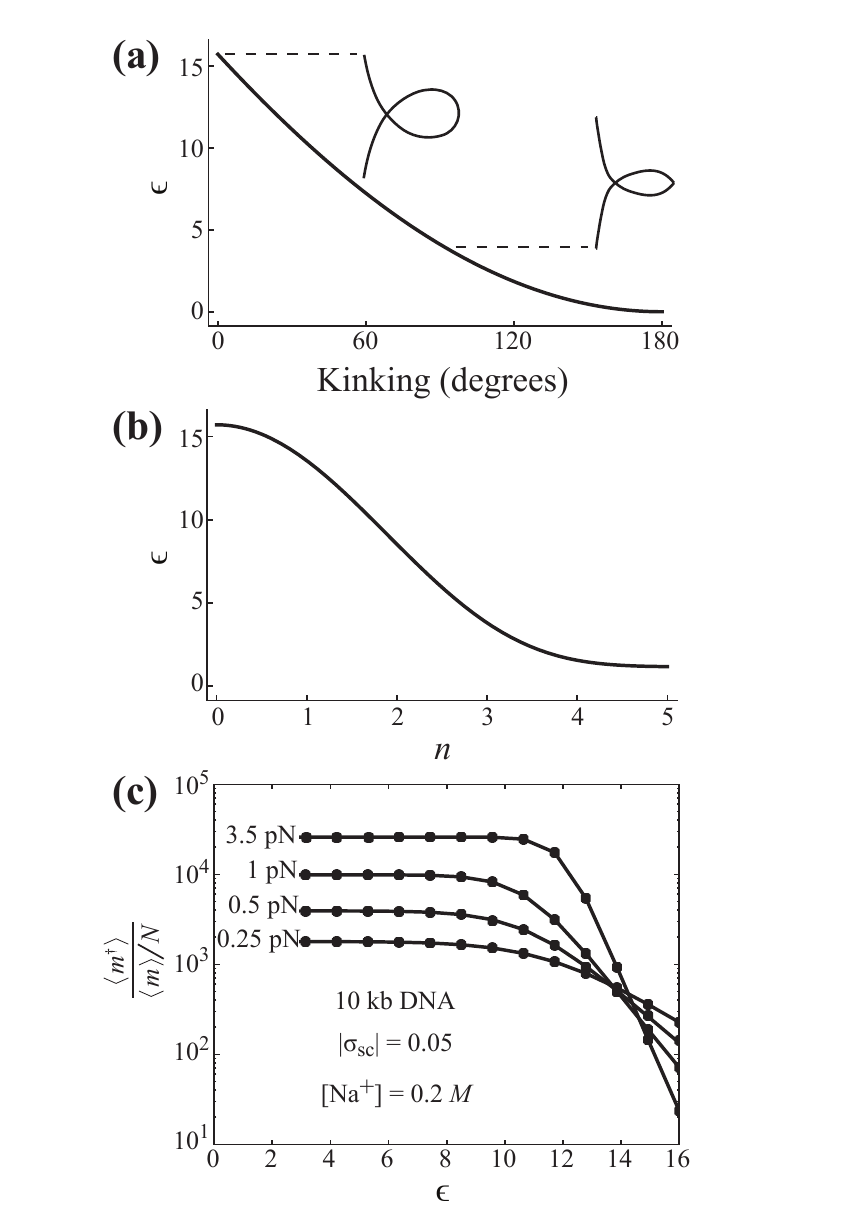}
    \caption{Model of end-loop kinking to estimate plectoneme pinning probabilities. (a) Energy scaling factor $\epsilon$ calculated for a planar end-loop as a function of kinking angle. (The interior apex angle and kinking angle are supplementary.) (b) Estimate of $\epsilon$ with defect size based on data in Fig.~2c. (c) Estimated enhancement, $\langle m_\epsilon^\dagger\rangle/(\langle m\rangle/N)$, of plectoneme occupancy at the defect relative to any other position along a $N=10$~kb topological domain of DNA at supercoiling density 0.05 and $0.2~M$ monovalent salt \cite{postow}. Points indicate independent calculations at each force and kinked end-loop energy.
    }
    \label{Fig4}
\end{figure}

To demonstrate this measurement approach, we tested torsionally constrained, 6~kb DNA molecules with $n=0$, 1, 2, 4, or 16 adjacent mismatches (Fig.~2). We positioned the mismatch site roughly 8$\%$ from one end of the DNA using a cassette based single-strand nicking template generated by PCR \cite{luzzietti} and simultaneously ligated a mismatch-containing oligonucleotide and handles for torsional constraint \cite{seol}. Upon supercoiling we observed re-buckling at the expected DNA extension $2\times8\%=16\%$ below the maximum. This confirms a single plectoneme formed and was pinned by the defect for all defect sizes, even a single mismatch ($n$=1). We do not observe re-buckling in DNA molecules with a centered defect, or in intact DNA ($n$=0). The re-buckling signal also disappears below the salt-force phase boundary demarcating the edge of the single-plectoneme regime (Fig.~\ref{Fig3}); at lower forces or ionic strengths, multiple plectonemes are expected \cite{marko}, and the pinned plectoneme may begin to exchange length with other plectonemes or diffuse \cite{vanloenhout}. All curves shown in Fig.~2 were collected at a fixed force of $f=3.6\pm0.4$~pN and 1~$M$ monovalent salt; these conditions favor a single plectoneme \cite{marko,vanloenhout}, which we find to be stably pinned at the mismatch.

\emph{Defects cause kinking of the pinned plectoneme end-loop} -- We focus on the initial plectoneme formation, which occurs at the mismatch whenever re-buckling is observed (Fig.~2). The abrupt extension drop, $\Delta X$, upon initial buckling decreases quadratically with defect size, $n$ (Fig.~2a,c). This is consistent with the expectation that the plectoneme end-loop decreases in size and is sharply bent at its apex to a degree that depends on the extent of the mismatch (Fig.~1). Crucially, defects reduce the critical buckling torque $\Gamma_{\rm c}$ occurring at linking number ${\rm Lk}^\dagger$ \cite{note1}.

To quantify the critical buckling torque $\Gamma_{\rm c}$ as a function of defect size, we define $\Gamma_{\rm c}$ relative to an internal reference torque, $\Gamma^*$, which is constant in the DNA after re-buckling and estimated using the analytic formula derived by Clauvelin, Audoly, and Neukirch \cite{clauvelin,mosconi} (Fig.~1): $\Gamma^*=21.0\pm0.5$~pN$\cdot$nm. This analysis also provides estimates of the structural parameters of the plectoneme \cite{clauvelin}: $\alpha=23.8\pm0.4^{\circ}$ is the ply angle and $r=1.7\pm0.1$~nm is its radius, consistent with a high degree of electrostatic screening \cite{stigter}.

For intact DNA ($n=0$) we assume that $\Gamma_{\rm c}-\Gamma^*$ is well approximated by the discrete torque drop $\Delta \Gamma_0$, which can be derived from the equilibrium buckling probability obtained from analysis of the extension fluctuations (Fig.~2a). Using measurement of $K=P_{unbuckled}/P_{buckled}$ from the relative probabilities of the unbuckled and buckled states (Fig.~2a), we estimate the free energy difference $\Delta F=-k_BT\ln{K}$ and note that torque is the partial derivative of free energy with respect to rotation angle, $2\pi {\rm Lk}$: $\Delta\Gamma=-k_BT/(2\pi)(\partial{\ln{K}}/\partial{{\rm Lk}})$. Our measurements at controlled ${\rm Lk}$ with all other differential quantities in the potential fixed permit straightforward evaluation of the partial derivative and model-free estimation of $\Delta\Gamma$. Under the assumption that $\Gamma_{\rm c}-\Gamma^*$ is proportional to the linking number difference, ${\rm Lk}^\dagger-{\rm Lk}^*$ (Fig.~1, Fig.~2d), we calculate $\Gamma_{\rm c}(n)\approx\Gamma^*+\Delta\Gamma_0({\rm Lk}^\dagger_n-{\rm Lk}^*_n)/ ({\rm Lk}^\dagger_0-{\rm Lk}^*_0)$.
Together, these estimates provide the critical torque over the range of abrupt buckling ($n<5$) (Fig.~2d). For larger defects ($n>5$) abrupt buckling is not observed. We conclude that for more than five adjacent mismatches there is no end-loop and abrupt buckling is replaced by a kinetic pathway without a transition barrier. Consistent with this, the $n=16$ curve (Fig.~2b) shows a continuous transition from the buckling point, ${\rm Lk}^*$; {\it i.e.}, the plectoneme forms once $\Gamma\approx\Gamma^*$.

For $n=16$, $\Delta\Gamma=0$. Surprisingly, $\Delta \Gamma$ for each of the smaller defects is nearly constant with an average value of $\Delta\Gamma=2.92\pm 0.08$~pN$\cdot$nm for $n=0$ to $n=4$. This indicates that the torque initially drops below $\Gamma^*$ but approaches $\Gamma^*$ with increasing ${\rm Lk}$. This undershoot is consistent with a more compact initial plectoneme structure due to the decreased bending energy of the end-loop.

\emph{Kinking lowers the end-loop energy} -- These measurements establish a relationship between end-loop energy and defect size. The end-loop energy can be related to geometry in a simple planar loop model \cite{sankararaman}. The size of the end-loop is obtained via minimizing the elastic energy,
\begin{equation*}
\beta\mathcal{E}_\gamma=\epsilon\frac{A}{\gamma}+\mu\frac{\gamma}{A},
\end{equation*}
where $\beta$ is the inverse thermal energy, $A$ is the bending persistence length, $\mu\equiv\beta A f$, and $\epsilon$ is a numerical prefactor related to the elastic energy of a loop of size $\gamma=A\sqrt{\epsilon/\mu}$. Notably, $\epsilon\approx16$ for a ``teardrop'' shape, and $\epsilon\approx4$ for a loop with its apex kinked by 90$^\circ$ (Fig.~4a) \cite{sankararaman}. Within the context of this model, the end-loop energy is related to the extension change $\Delta X(n)$ (Fig.~2c) through $\gamma(n)$. The data in Fig.~2 correspond to plectonemes with end-loops ranging from zero kinking at $n=0$ to a collapsed loop for large defects ($n>5$), which we assume to be kinked to a limiting interior apex angle of $2\alpha$. By fixing these limits and fitting to $\Delta{X(n)}$ (Fig.~2c), we estimated $\epsilon$ vs $n$ for our experimental data set (Fig~4b).

\emph{End-loop kinking model predicts a preference for pinned plectonemes} -- Outside the regime where there is a single pinned plectoneme (Fig. 3) we lose the re-buckling signal and our experimental method cannot unambiguously determine the position or number of plectonemes: Plectoneme pinning at the mismatch becomes probabilistic rather than deterministic \cite{vanloenhout}. To obtain these probabilities, we turn to theory, employing the energy-scaling parameter $\epsilon$ obtained from experiments (Fig.~4b).

Building on a detailed mesoscopic model described previously \cite{marko}, we consider a DNA molecule of contour length $L$ with defect positioned at $L$'. We sum over all possible sizes and numbers of plectonemes to construct a canonical partition function that includes $m\in[0,1,2\dots]$ plectonemes with non-kinked loops and the $m^\dagger\in[0,1]$ pinned plectoneme with its end-loop kinked by the defect.

In calculating the partition sum, both energetic and entropic contributions are accounted for in the fixed force and fixed linking number ensemble. The DNA is partitioned into a force-extended fraction and plectonemic fraction. The extended fraction includes the free energy of twisting as well as the total energy associated with force-extension and lateral fluctuations. The plectonemic fraction includes twist energy, superhelical bending, electrostatic repulsion, and the elastic energy contribution of plectoneme end-loops ($\epsilon<16$ at a defect and $\epsilon_0=16$ elsewhere); in addition, we explicitly include contributions to the free energy of fluctuations in the plectoneme \cite{brahmachari}, and entropy correction factors to account for plectonemes that are mobile and exchange length. We minimize free energy to calculate the equilibrium values of the number of mobile plectoneme domains $\langle  m\rangle$, and occupancy of the pinned plectoneme ($0\leq\langle m^\dagger\rangle\leq1$) at the defect. We will describe the details of this calculation and its range of predictive results in a forthcoming article.

In essence, the decrease in bending energy at the defect competes with the loss of entropy associated with a defect-pinned plectoneme. The pinned plectoneme is expected to be present in the fluctuating system with a fractional occupancy that approaches 1 only in the limit of high tension and ionic screening (Fig.~3). Although at lower force and ionic strength, multiple plectonemes may be present and the defect causes only a small change to the global probability of plectoneme formation, a plectoneme pinned at the defect represents a large local change, which we can now calculate, even if its fractional occupancy is $\ll1$.

To assess whether supercoiling could influence DNA damage sensing and repair, we focus on conditions most relevant to DNA in the bacterial cell \cite{postow} and calculate 
$\langle m\rangle$ and
$\langle m^\dagger\rangle$ for an average $10$~kb topological domain at supercoiling density 0.05 and $0.2~M$ monovalent salt. Whereas the $\langle m\rangle$ diffusing plectonemes are randomly distributed across the $N=10^4$ base pairs, the pinned plectoneme with fractional occupancy $\langle m^\dagger\rangle$ occurs only at the defect ($\sim1$ base pair). We therefore compare the ratio of probabilities, $\langle m^\dagger\rangle/(\langle m\rangle/N)$, yielding the relative enhancement of plectoneme occupancy at the defect as a function of $\epsilon$ (Fig.~4c). We restrict our attention to small defects ($n<2$) and note that the values ($8<\epsilon<16$) are expected to cover a range of different defects on the scale of 1~bp, including mismatches, abasic sites, and other lesions \cite{sharma,yang}. Relative to a random position on the DNA, kinking generally enhances plectoneme occupancy at the defect by a factor that depends on force and the end-loop kinking energy (Fig.~4c). The experimental phase diagram (Fig.~3) is broadly consistent with the calculated force dependence of defect-enhanced buckling in Fig.~4. At 3.5~pN $\langle m^\dagger\rangle/\langle m\rangle>1$, consistent with the observation of re-buckling at this force (Fig.~3). At the low force of 0.25~pN and for end-loop energies corresponding to a single base-pair mismatch (Fig.~4b; $\epsilon\approx13.5$ for $n=1$), physiologically relevant supercoiling promotes buckling at the defect by a factor  $\sim10^2-10^3$ (Fig.~4c).

\emph{Repair of mismatches and damaged bases begins with sensing} -- The single-molecule data and theory presented here establishes that supercoiling localizes mismatches to the tips of kinked plectoneme end-loops. From a biological perspective, our data suggest a role for supercoiling in defect sensing and repair. Since plectonemes project outward from the dense bacterial nucleoid, presentation of defects at plectoneme tips could promote access to lesion sites and accelerate the diffusive search by proteins that bind and initiate repair \cite{gowers,lomholt,irovalieva}. Furthermore, binding at the lesion is accompanied by DNA kinking and stabilization of base-flipped configurations \cite{obmolova,lamers,hosfield}. Since bending stress at the plectoneme end-loop favors these protein-bound structural states, tighter protein binding is expected from an energetic standpoint. The hundred-fold or greater relative enhancement of plectoneme occupancy at a mismatch under physiological supercoiling conditions could therefore promote mismatch detection by a similar or possibly larger factor.

This work was supported by in part by the Intramural Research Program of the National Heart, Lung, and Blood Institute, National Institutes of Health. Work at NU was supported by the NIH through grants R01-GM105847, U54-CA193419 (CR-PS-OC) and a subcontract to grant U54-DK107980, and by the NSF through grants MCB-1022117 and DMR-1206868.


\makeatother
\pagebreak
\onecolumngrid
\section*{Supporting Information}\subsection*{Supercoiling DNA locates mismatches}
\setcounter{equation}{0}
\setcounter{figure}{0}
\setcounter{table}{0}
\setcounter{page}{1}
\makeatletter
\renewcommand{\theequation}{S\arabic{equation}}
\renewcommand{\thefigure}{S\arabic{figure}}

{\it DNA preparation methods} -- Using the Bam-HI restriction site, we inserted the forward sequence GATCGCTGAGGCGCAGCTTCCGACTGCAGCCTGACGCCAGGGCTGAGGT after position 198 in the pET-28b plasmid. This introduced both a central Pst-I restriction site (CTGCAG) and two sites for the nicking enzyme Nt.BbvCI. Successful insertion also eliminated the Bam-HI site to allow for clean selection of the recombinant DNA prior to transfection. We amplified a region of this plasmid DNA with Phusion polymerase using the forward primer -GAACCATCACCCTAATCAAG and reverse primer -GAACAACACTCAACCCTATC. These were prepended by the overhang sequences GCTGGGTCTCGCAAC- and  GCTGGGTCTCGACCA-, respectively, to introduce nonpalendromic Bsa-I cleavage sites \cite{seol}. After purification of the PCR product, we simultaneously nicked with Nt.BbvCI and cut with Bsa-I. We then annealed a 10-fold excess of the sequence GGCGCAGCTTCCGACTGCAGCCTGACGCCAGGGCTGA to competitively hybridize with the excised DNA fragment. Upon second purification, this produced DNA with an internal 37-nt gap and unique 4-nt sticky overhangs. We ligated 0.5~kb handles containing the complementary 4-nt overhangs and functionalized with either multiple biotin or digoxigenin moieties \cite{seol} and each of the following sequences to introduce a 1, 2, 4, or 16~bp mismatch (underscore):
\begin{itemize}
\item[]
$n=1$:  ~~/5'Phos/TCAGCCCTGGCGTCAGGC\underline{A}GCAGTCGGAAGCTGCGCC
\item[]
$n=2$:  ~~/5'Phos/TCAGCCCTGGCGTCAGGC\underline{AC}CAGTCGGAAGCTGCGCC
\item[]
$n=4$:  ~~/5'Phos/TCAGCCCTGGCGTCAGG\underline{GACG}AGTCGGAAGCTGCGCC
\item[]
$n=16$:~~/5'Phos/TCAGCCCTGG\underline{GCAGTCCGACGTCAGC}GAAGCTGCGCC
\end{itemize}
Successful insertion of the mismatch-containing sequence destroys the Pst-I site. This allowed us to finish the product with Pst-I digestion, ensuring a pure sample for testing in magnetic tweezers. Intact DNA was prepared similarly but without Nt.BbvCI or Pst-I.\\

{\it Isolation and testing of single DNA molecules using magnetic tweezers} -- Each DNA molecule ($2\pm0.2~\mu$m contour length) was isolated and torsionally constrained between an antibody-coated glass surface and a $\sim1~\mu$m streptavidin-coated paramagnetic bead (Dynal) using multiple biotin or digoxigenin moieties at either extremity. A magnetic field gradient, controlled by the proximity of permanent magnets, pulled the bead upward and extended the DNA. At constant force, magnet rotation synchronously rotated the bead to add turns to the DNA, thereby controlling its topology. We performed all measurements in Fig.~2 in $10$~m$M$ Tris buffer, pH 7.5 with 1~$M$ NaCl and 0.1$\%$ Tween-20. For the data in Fig.~3, we used the same buffer conditions but varied the salt concentration.\\

\begin{figure}
    \centering
    \includegraphics[width=86mm]
    {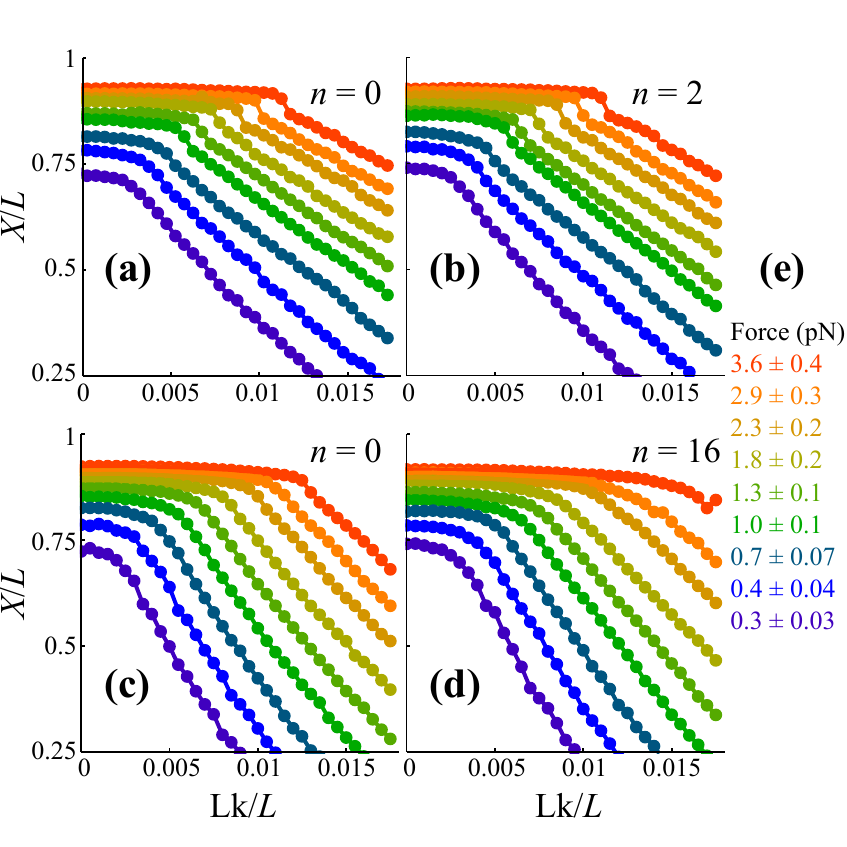}
    \caption{Plots of normalized mean extension, $X/L$, vs excess linking density, ${\rm Lk}/L$, for $L=2~\mu$m DNA molecules. Measurements on (a) intact DNA and (b) DNA with $n=2$ mismatches in $1~M$ monovalent salt show no apparent differences. Measurements on (c) intact DNA and (d) DNA with $n=16$ mismatches in $0.16~M$ monovalent salt show torsional compliance is enhanced by the defect. The values of applied forces are tabulated in (e). Errors at each measurement are smaller than the point size.}
    \label{FigS1}
\end{figure}

{\it Comparison to traditional magnetic tweezers measurements} --
In the absence of the re-buckling signal (Fig.~1), traditional magnetic tweezers measurements comparing DNA molecules with $n=0$ and $n=2$ mismatches show no apparent differences (Fig.~\ref{FigS1}(a,b)). In contrast, analysis of the densely sampled data in Fig.~2 shows the defect causes buckling to occur at a lower critical torque, $\Gamma_{\rm c}(n)$, which, ignoring small differences in torsional elasticity, is assumed to linearly increase from $\Gamma^*$ in proportion to ${({\rm Lk}_n^\dagger-{\rm Lk}_n^*)}$ (Fig.~1, Fig.~2d). Compensating trends account for differences in $\Gamma_{\rm c}(n=0)$ and $\Gamma_{\rm c}(n=2)$ even though ${\rm Lk}_0^\dagger$ and  ${\rm Lk}_2^\dagger$ are similar: As (${\rm Lk}_n^\dagger-{\rm Lk}_n^*$) is reduced by \emph{bending} compliance, \emph{torsional} compliance of the defect causes ${\rm Lk}_n^*$ to increase. The enhanced torsional compliance is noticeable for a large defect (Fig.~\ref{FigS1}(c,d)). These trends are repeatable across different constant forces (tabulated in Fig.~\ref{FigS1}e).

\end{document}